\documentclass[cameraready]{Interspeech}

\title{DAST: A \textbf{D}ual-Stream Voice Anonymization \textbf{A}ttacker with \textbf{S}taged \textbf{T}raining}
\author[affiliation={1}, orcid=0009-0007-7464-4959]{Ridwan}{Arefeen}
\author[affiliation={2}, orcid=0000-0002-6645-6524, correspondingauthor]{Xiaoxiao}{Miao}
\author[affiliation={1}, orcid=0000-0003-3410-8354]{Rong}{Tong}
\author[affiliation={3}]{Aik Beng}{Ng}
\author[affiliation={3}]{Simon}{See}
\author[affiliation={3}]{Timothy}{Liu}

\address{
    $^1$ Singapore Institute of Technology, Singapore \\
    $^2$ Duke Kunshan University, China \\
    $^3$ NVIDIA AI Technology Centre, Singapore
}
\email{2403754@sit.singaporetech.edu.sg, xiaoxiao.miao@dukekunshan.edu.cn, tong.rong@singaporetech.edu.sg}


\usepackage{comment}
\usepackage{todonotes}
\begin{document}

\maketitle

\begin{abstract}
Voice anonymization masks vocal traits while preserving linguistic content, which may still leak speaker-specific patterns. To assess and strengthen privacy evaluation, we propose a dual-stream attacker that fuses spectral and self-supervised learning features via parallel encoders with a three-stage training strategy. Stage I establishes foundational speaker-discriminative representations. Stage II leverages the shared identity-transformation characteristics of voice conversion and anonymization, exposing the model to diverse converted speech to build cross-system robustness. Stage III provides lightweight adaptation to target anonymized data. 
Results on the VoicePrivacy Attacker Challenge (VPAC) dataset demonstrate that Stage II is the primary driver of generalization, enabling strong attacking performance on unseen anonymization datasets. With Stage III, fine-tuning on only 10\% of the target anonymization dataset surpasses current state-of-the-art attackers in terms of EER.\footnote{Full code and pretrained models can be found at \textbf{https://github.com/monkeyDarefeen/DAST}} .
\end{abstract}

\section{Introduction}

Speech data inherently encodes a wide range of privacy-sensitive attributes, including speaker identity, age, gender, health condition, and socio-economic status. Unauthorized disclosure can lead to serious risks such as identity theft, surveillance misuse, and other privacy violations \cite{Nautsch-PreservingPrivacySpeech-CSL-2019}. Voice privacy \cite{tomashenko2020introducing} has therefore emerged as a critical research objective, commonly framed as a game between defenders and attackers. 

The VoicePrivacy Challenge~(VPC) series~\cite{tomashenko2020introducing,tomashenkovoiceprivacy,Tomashenko2021CSl, tomashenko2024voiceprivacy, tomashenko2026third} has provided standardized benchmarks and significantly advanced voice anonymization research. However, privacy protection reported under fixed attacker models may be overestimated, as such results can reflect limited attacker generalization rather than truly robust anonymization~\cite{panariello2025risks}. The VoicePrivacy Attacker Challenge~(VPAC)~\cite{tomashenko2024first, tomashenko2025first} addresses this gap by tasking participants with developing stronger attacker systems, specifically, automatic speaker verification~(ASV) systems that determine whether an anonymized enrollment--trial pair originates from the same speaker\footnote{VPAC adopts a \textit{semi-informed} attacker model in which users anonymize enrollment utterances, while attackers have access to the anonymization system and attempt to re-identify the original speaker behind each anonymized trial utterance.} against seven anonymization systems.

From an \emph{architectural} perspective, existing approaches span:
(i)~\emph{data augmentation} via spectrogram resizing~\cite{10890291} and segment rearrangement~\cite{segreconcat};
(ii)~\emph{feature representations} using self-supervised learning (SSL) features~\cite{10890291,lyu2025fast} or multimodal inputs~\cite{aloradi2025voxattack};
(iii)~\emph{architecture adaptation} through LoRA-adapted ResNet34~\cite{lyu2025fast} and fine-tuned TitaNet-Large~\cite{mawalim2025fine};
(iv)~\emph{backend refinement} with PLDA scoring~\cite{zhang2025attacking} and score normalization~\cite{xinyuan2025hltcoe}.
From a \emph{training strategy} perspective, most attackers train exclusively on speech anonymized by the target system, or combine original speech with the corresponding anonymized data~\cite{zhang2025attacking}, or manually select speech from a subset of systems through trial and error~\cite{mawalim2025fine}. The latest work~\cite{aloradi2025voxattack} trains on speech from all seven systems simultaneously, yielding better results than single-system training for some target systems but not all, revealing that cross-system generalization remains challenging.

We argue that architecture and training strategy should be addressed jointly: a single feature view cannot capture the full range of residual speaker cues, yet even a multi-view architecture will overfit if trained on only one target system. To this end, we propose a \textbf{D}ual-stream voice anonymization \textbf{A}ttacker with three-\textbf{S}tage \textbf{T}raining~(DAST). All three stages share a dual-stream model that utilizes SSL features and spectral~(Fbank) features, each passed to separate ECAPA-TDNN frame encoders and fused at the mid-level to capture complementary high-level and fine-grained speaker characteristics. The three-stage design follows a deliberate curriculum: in Stage~I, the model is trained on original, unprocessed speech to establish \emph{what} to look for (speaker-discriminative cues); in Stage~II, since voice anonymization and voice conversion share the same underlying mechanism, transforming speaker identity from a common source, the model is trained on large-scale, diverse voice-converted data to learn \emph{where} to find them under distortion (anonymization-invariant representations); and in Stage~III, fine-tuning on target anonymized data sharpens the model for \emph{how} a specific system distorts them.

Experiments demonstrate that: (1)~Stage~II is the primary driver of generalization: the model achieves strong cross-system performance without any target-specific adaptation; and (2)~ablation studies on Stage~III show that the proposed model adapts rapidly to unseen anonymization systems: fine-tuning with 10\% of target anonymization dataset already surpasses the current best attacker, with full-dataset fine-tuning yielding further improvements, and consistently outperforming existing attackers across all anonymization systems under the VPAC protocol.

\section{Proposed Method}
\label{sec:method}
\begin{figure}[h] 
    \centering
    \includegraphics[width=1\columnwidth, trim=18 10 18 18, clip]{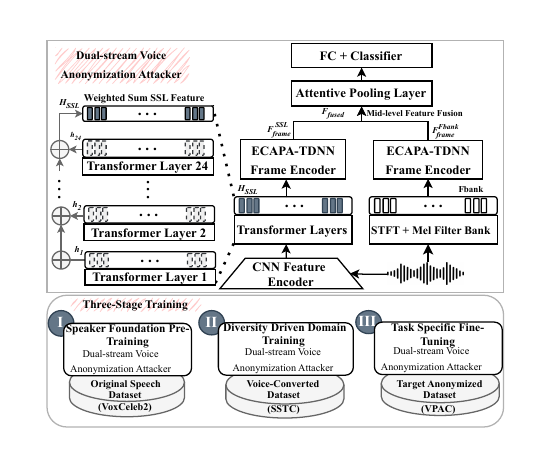}
    \vspace{-8mm}
    \caption{Proposed Method: (top) dual-stream architecture with mid-level feature fusion (bottom) three stage training strategy}
    \label{fig:architecture}
\end{figure}

Figure~\ref{fig:architecture} illustrates the proposed system: (top) the dual-stream architecture and (bottom) the three-stage training pipeline. We first describe the architecture design, then detail the staged training strategy.

\subsection{Dual-Stream Architecture}
Fusing handcrafted spectral and learned representations has proven effective in conventional ASV tasks~\cite{chen2022largescaleselfsupervisedspeechrepresentation}. Building on this, we introduce a dual-stream architecture for the attacker ASV task that processes two complementary input streams: Mel-filterbank~(Fbank) features and self-supervised learning~(SSL) features.

\noindent\textbf{Feature extraction.}
The spectral stream computes Fbank features via short-time Fourier transform followed by a Mel filterbank, capturing fine-grained acoustic information such as formant structure and spectral envelope. The SSL stream passes raw waveforms through a frozen, pre-trained WavLM \cite{9814838} model and extracts hidden states from all transformer layers. To aggregate complementary multi-scale and multi-level information, we apply a learnable weighted sum, $H_{\text{SSL}} = \sum_{l=1}^{L} w_l \cdot h_l$, where $h_l$ is the hidden state of the $l$-th transformer layer and $w_l$ is a learnable scalar weight. This mechanism adaptively balances low-level acoustic cues with higher-level phonetic and contextual representations.

\noindent\textbf{Parallel frame encoding.}
Unlike early-fusion approaches \cite{chen2022largescaleselfsupervisedspeechrepresentation} that merge heterogeneous features before encoding, we process each stream through a dedicated ECAPA-TDNN frame encoder with independent parameters, \emph{i.e.}, $F_{\text{frame}}^{\text{SSL}} = \text{ECAPA}_{\theta_1}(H_{\text{SSL}})$ and $F_{\text{frame}}^{\text{Fbank}} = \text{ECAPA}_{\theta_2}(X_{\text{Fbank}})$, where $\theta_1$, $\theta_2$ are separate parameter sets. Each encoder extracts multi-scale temporal features via stacked convolutional blocks with squeeze-and-excitation modules. Isolating the two streams during encoding allows each encoder to specialize in its respective feature domain.

\noindent\textbf{Mid-level fusion.}
After frame encoding, the two streams are fused via element-wise multiplication, $F_{\text{fused}} = F_{\text{frame}}^{\text{SSL}} \odot F_{\text{frame}}^{\text{Fbank}}$, where $\odot$ denotes the Hadamard product. This formulation allows the SSL stream to dynamically recalibrate the spectral stream to emphasize speaker-discriminative patterns while suppressing conversion-induced artifacts. We adopt mid-level fusion, rather than the early fusion commonly adopted in conventional ASV \cite{chen2022largescaleselfsupervisedspeechrepresentation},  based on the assumption that in anonymized speech, conversion artifacts are entangled with speaker characteristics at different feature levels, making it difficult for a single encoder to disentangle both simultaneously. By fusing after independent encoding, each stream has already partially separated speaker-relevant from artifact-related patterns, enabling more effective calibration. 

\noindent\textbf{Pooling and classification.}
The fused frame-level features are aggregated into a fixed-dimensional utterance-level embedding via Attentive Statistics Pooling~(ASP), which computes attention-weighted mean and standard deviation over the temporal dimension. The embedding is then projected through a fully connected layer and optimized with AAM-Softmax \cite{wang2018additive}. 
%This dual-stream model is used to do the following staged training.
This dual-stream architecture serves as the shared backbone across all three training stages described below.

\subsection{Three-Stage Training Strategy}
\label{sec_three_t_s}
\noindent\textbf{Stage~I: Speaker Foundation Pre-training.}
Before mitigating the effects of anonymization, the model must first acquire strong speaker-discriminative representations under clean acoustic conditions. To achieve this, we train the dual-stream architecture on unprocessed speech data. During this stage, the WavLM model is kept frozen to preserve its pre-trained representations; consequently, optimization is localized entirely to the downstream components: the learnable layer weights, the dual ECAPA-TDNN frame encoders, the feature fusion mechanism, and the classifier. This allows the downstream components to learn how to extract and combine speaker-discriminative cues from the two feature streams without corrupting the general-purpose SSL representations that will be essential for generalization in later stages.

\noindent\textbf{Stage~II: Diversity-Driven Domain Training.}
With speaker-discriminative representations established, Stage~II aims to make them robust to anonymization-induced distortions. The key observation is that voice anonymization and voice conversion share the same underlying mechanism \cite{cai2024privacy,das2024comparing, miao2023language,yao2025easy}: both transform speaker identity from a common source speech signal using neural vocoders or voice conversion models. They differ only in intent (privacy vs.\ style transfer), not in the types of distortions: both alter pitch, formant structure, and speaker embedding characteristics while preserving linguistic content. A model trained to maintain speaker discrimination across diverse voice conversion systems should therefore inherently learn to handle anonymization-induced distortions as well.

Following this reasoning, we train on diverse voice converted data generated by different voice conversion (VC) approaches. The diversity of VC data is essential: each introduces distinct artifact patterns (spectral envelope distortion, prosody alteration, vocoder-specific noise, etc.), forcing the model to rely on speaker cues that persist across various transformations rather than overfitting to any single system's artifacts. Within the dual-stream architecture, the SSL stream learns gating masks that are invariant to synthetic artifacts, while the spectral stream retains sensitivity to residual acoustic speaker cues. Together, this decouples identity-specific features from transformation-induced noise.

\noindent\textbf{Stage~III: Task-Specific Adaptation.}
Although Stage~II has already learned anonymization-invariant representations, each unseen anonymization system has its own characteristic distortion pattern that the model has never observed. Stage~III bridges this gap through fine-tuning on anonymized speech from the target system in the VPAC challenge~\cite{tomashenko2024first, tomashenko2025first}. Rather than relying on the full amount of target-specific anonymized data, we hypothesize that this refinement is highly sample-efficient: a small fraction of the available data should suffice to achieve strong attack performance on the target system. This significantly reduces the computational cost of adaptation and enables rapid privacy evaluation of newly proposed anonymization systems, an attacker need only process a small number of utterances through the target system to mount an effective attack, providing a more realistic and stringent test of anonymization robustness.

\section{Experiments}

To evaluate the proposed DAST attacker model, we conduct systematic experiments within the framework of the VPAC~\cite{tomashenko2024first, tomashenko2025first}. We first validate the dual-stream architecture with mid-level fusion by comparing it against single-stream baselines using either Fbank or weighted-sum SSL features alone, as well as alternative fusion strategies (early fusion vs.\ mid-level fusion). All models in this phase are trained from scratch on each specific anonymized dataset from VPAC separately (\emph{i.e.}, Stage~III only), isolating the architectural contribution from the training strategy.
With the dual-stream architecture validated, we adopt it as the backbone and progressively evaluate the three-stage training strategy. 

\subsection{Datasets and Evaluation Metrics}

To support the three-stage training curriculum, we utilize three distinct corpora, each aligned with a specific training objective as described in Section~\ref{sec:method}. The Stage~I model is trained on VoxCeleb2~\cite{chung2018voxceleb2} over 1 million utterances, 5{,}944 speakers, a large-scale in-the-wild unprocessed speaker dataset. The Stage~II model uses the training set of the Source Speaker Tracing Challenge~(SSTC)~\cite{SSTC}, where LibriSpeech~\cite{panayotov2015librispeech} (1{,}172 speakers) serves as source speech and VoxCeleb~\cite{chung2018voxceleb2} as target speech, processed by eight different voice conversion systems (AGAIN-VC~\cite{againvc}, FreeVC~\cite{freevc}, MediumVC~\cite{mediumvc}, StyleTTS~\cite{styletts}, TriAAN-VC~\cite{triaan}, VQMIVC~\cite{vqmivc}, KNN-VC~\cite{knnvc}, and Sig-VC~\cite{zhang2022sig}), totaling over 2.6 million utterances (327,600 * 8, ${\sim}$9k hours). 

In Stage~III, we strictly follow the VPAC protocol and fine-tune on the anonymized VPAC data, i.e., \textit{LibriSpeech train-clean-360} (921 speakers, 104{,}014 utterances)~\cite{panayotov2015librispeech}, where the data are anonymized separately by seven distinct anonymization systems (B3, B4, B5, T8-5, T10-2, T12-5, and T25-1). Note that the eight voice conversion systems used in SSTC are entirely disjoint from the seven anonymization systems in the VPAC evaluation, ensuring that any cross-system performance gain reflects genuine generalization rather than data leakage.

Beyond full fine-tuning on the anonymized \textit{LibriSpeech train-clean-360} set, we further investigate a lightweight Stage~III adaptation by sampling a fixed percentage of utterances per speaker (5\%--50\%) while retaining all 921 speakers, thereby preserving speaker diversity and proportionally reducing the amount of training data per speaker.\footnote{We also experimented with random sampling without enforcing speaker diversity preservation and observed that retaining all speakers yields better performance. Therefore, we only report the results obtained with speaker-diversity-preserved sampling in this paper.}

Following VPAC evaluation protocol, we report EER, defined as the average of female and male equal error rates on the VPAC test sets for each system, to ensure fair comparison with prior work. A lower EER indicates a stronger attack.

\subsection{Experimental Setup}

\begin{figure*}[h] 
     \centering
     \includegraphics[width=1\textwidth]{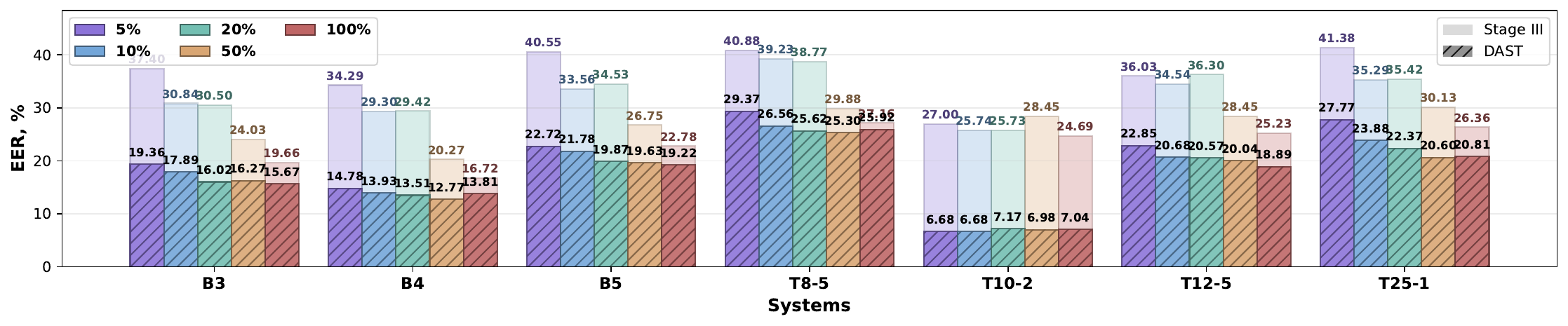}
     \vspace{-7mm}
     \caption{Comparison of the DAST and Stage III EER, \% on VPAC Benchmark for low-resource adaptation.}
     \label{fig:low_resource_figure}
 \end{figure*}

The spectral stream uses 80-dimensional Fbank features. The SSL stream extracts 1{,}024-dimensional features from WavLM-Large\footnote{\url{https://huggingface.co/microsoft/wavlm-large}}. SpecAug data augmentation \cite{Park_2019}, which masks blocks of frequency channels and time steps, is applied. All experiments share the same ECAPA-TDNN backbone with 1{,}024 channels in the convolutional frame layers, implemented via SpeechBrain\footnote{\url{https://speechbrain.readthedocs.io/}}. Speaker embeddings are extracted from the final layer, and similarity scoring is performed using cosine distance between averaged enrollment and test embeddings.
When transitioning to Stage~III, we replace the final classification layer to match the number of speakers in the target anonymized dataset (921 speakers).  
The initial learning rate for all stages is set to $10^{-3}$ and scheduled using the cyclical learning rate (CLR) policy~\cite{smith2017cyclical}.
AdamW is used as the default optimizer, except in Stage~II where the Muon optimizer\footnote{To effectively handle the massive data utilization in Stage II, we utilize the Muon optimizer, which achieves faster convergence and better generalization by maintaining weight matrix orthogonality during training, comprared to AdamW.} \cite{jordan2024muon} is adopted. Additionally, Stages~I and II use a batch size of 64 with 4 gradient accumulation steps, whereas Stage~III reduces the batch size to 32 with no gradient accumulation. 
Dropout is set to 0.0 in Stage~I and increased to 0.3 in Stages~II and III to enhance robustness. 
Training lasts for 1 epoch in Stage~I, 2 epochs in Stage~II, and varies in Stage~III depending on the data ratio: 2 epochs when using 100\% of the data, 4 epochs for 50\%, and 6 epochs for low-resource settings (5--20\%) to compensate for reduced supervision.
All experiments were conducted using a single NVIDIA H200 GPU\footnote{The code and pretrianed models will be released upon acceptance to facilitate reproducibility}.

\begin{table}[t]
\centering
\footnotesize
\caption{Comparison of feature representations and fusion strategies, trained from scratch on each anonymization system (EER, \% $\downarrow$ ).}
\vspace{-2mm}
\label{tab:feature_comparison}
\setlength{\tabcolsep}{3pt}
\begin{tabular}{@{}l ccccccc@{}}
\toprule
& \textbf{B3} & \textbf{B4} & \textbf{B5} & \textbf{T8-5} & \textbf{T10-2} & \textbf{T12-5} & \textbf{T25-1} \\
\midrule
Fbank     & 27.32 & 30.26 & 34.34 & 41.28 & 40.36 & 42.75 & 42.13 \\
SSL       & 19.67 & 18.46 & 22.92 & 34.42 &    \textbf{23.02}   & 25.56 & 28.03 \\
Raw-level & 19.75 & 16.88 & 23.49 & 30.02 &   23.19    & 26.15 & 27.23 \\
Mid-level & \textbf{19.66} & \textbf{16.72} & \textbf{22.78} & \textbf{27.16} & 24.69 & \textbf{25.23} & \textbf{26.36} \\
\bottomrule
\end{tabular}
%\vspace{-3mm}
\end{table}

\begin{table}[t]
\centering
\footnotesize
\caption{Comparison of different training strategies (EER, \%). \checkmark\ indicates the stage is included.}
\label{tab:results}
\vspace{-2mm}
\setlength{\tabcolsep}{3pt}
\begin{tabular}{@{}p{0.4cm}p{0.4cm}p{0.4cm} ccccccc@{}}
\toprule
\textbf{I} & \textbf{II} & \textbf{III} & \textbf{B3} & \textbf{B4} & \textbf{B5} & \textbf{T8-5} & \textbf{T10-2} & \textbf{T12-5} & \textbf{T25-1} \\ \midrule
 \checkmark & &  & 43.25 & 45.55 & 48.86 & 44.36 & 34.39 & 49.41 & 49.57 \\
  & \checkmark &  & 23.32 & 18.56 & 28.66 & 31.81 & 8.53 & 29.37 & 32.71 \\
 & & \checkmark & 19.66 & 16.72 & 22.78 & 27.16 & 24.69 & 25.23 & 26.36 \\
 \checkmark & \checkmark & & 21.44 & 18.11 & 27.31 & 30.49 & 8.30 & 27.84 & 29.91 \\
 \checkmark & & \checkmark &  19.21 & 16.26 & 24.28 & 28.70 & 30.11 & 25.05 & 25.65 \\
\checkmark & \checkmark & \checkmark & 
\textbf{15.67} & \textbf{13.81} & \textbf{19.22} & \textbf{25.92} & \textbf{7.04} & \textbf{18.89} & \textbf{20.81} \\ 

\bottomrule
\end{tabular}
\vspace{-5pt}
\end{table}

\begin{table}[t]
\centering
\footnotesize
\caption{Comparison with existing systems (EER, \% ).}
\label{tab:comparison}
\setlength{\tabcolsep}{3pt}
\begin{tabular}{@{}lccccccc@{}}
\toprule
\textbf{Method} & \textbf{B3} & \textbf{B4} & \textbf{B5} & \textbf{T8-5} & \textbf{T10-2} & \textbf{T12-5} & \textbf{T25-1} \\ \midrule
VPAC-base & 27.32 & 30.26 & 34.34 & 41.28 & 40.36 & 42.75 & 42.13 \\
VPAC-Top1~\cite{lyu2025fast} & 20.51 & 19.56 & 25.51 & 26.39 & 22.46 & 25.63 & 27.77 \\
VoxAttack~\cite{aloradi2025voxattack} & 19.90 & 18.40 & 24.30 & 28.70 & 25.10 & 30.30 & 24.50 \\
\midrule
DAST (10\%) & 17.89 & 13.93 & 21.78 & 26.56 & 6.68 & 20.68 & 23.88 \\
\textbf{DAST (100\%)} & \textbf{15.67} & \textbf{13.81} & \textbf{19.22} & \textbf{25.92} & \textbf{7.04} & \textbf{18.89} & \textbf{20.81} \\
\bottomrule
\end{tabular}
\vspace{-5pt}
\end{table}
\vspace{-5pt}
\subsection{Experimental results}
\subsubsection{Results on dual-stream architecture}
Table~\ref{tab:feature_comparison} compares four configurations: single-stream Fbank, single-stream weighted-sum SSL, and dual-stream fusion at either the raw feature level or the mid-level. The observations are: the SSL stream alone substantially outperforms the Fbank stream across all systems, confirming the strength of pre-trained self-supervised representations for speaker discrimination under anonymization. Second, fusing the two streams consistently improves over either single-stream baseline, regardless of the fusion strategy, demonstrating that spectral and SSL features provide genuinely complementary information. Third, among the two fusion strategies, the proposed mid-level fusion achieves the best performance across most anonymization systems, supporting our argument that fusing after independent encoding, where each stream has already separated speaker-relevant from artifact-related patterns, enables more effective feature interaction than raw-level fusion. We thus take this dual-stream mid-level fusion architecture as the backbone and progressively evaluate the three-stage training strategy. 

\subsubsection{Results on three-stage training strategy}
\label{result_on_tst}

Table~\ref{tab:results} presents the ablation results across the training stages. The model trained only on unprocessed data (Stage~I) cannot reliably reidentify the source speaker, resulting in very high EERs of over 40\% across all systems. Training on large-scale domain-related VC data (Stage~II) reduces the EER by more than at least 12\% in absolute terms for all anonymization datasets, particularly for T10-2, the likely reason is that the Stage~II SSTC data contain VC approaches similar to those used in T10-2. Stage~III, which is trained on matched anonymized datasets, achieves much lower EERs. 

We then move to the progressive training. The Stage~I+II model, without any target-specific adaptation, achieves lower EERs than either Stage~I or Stage~II alone and reaches performance comparable to Stage~III, revealing that diversity-driven domain training alone can provide strong cross-system robustness.
Unlike the Stage~I+II model, simply using Stage~I pre-training as the starting point and applying Stage~III fine-tuning achieves better results for some systems but worse results for others compared to Stage~III training alone. This indicates that Stage~I pre-training alone does not consistently provide a beneficial initialization for adaptation to anonymized data. Finally, adopting all stages sequentially (last row) yields the best attacking performance.

\subsubsection{Low-Resource Adaptation}
As discussed in Section~\ref{sec_three_t_s}, we expect that DAST with the three-stage training is highly sample-efficient, requiring only a small fraction of target data to achieve strong attack performance. Figure~\ref{fig:low_resource_figure} confirms this: as the amount of training data decreases, the EER of the three-stage DAST increases much more slowly than that of models trained with Stage~III only using the same proportion of data.

\subsubsection{Comparison with Existing Systems}

Table~\ref{tab:comparison} compares DAST against the VPAC baseline, the challenge Top-1 attacker, and the state-of-the-art VoxAttack~\cite{aloradi2025voxattack}. DAST (100\%) achieves the lowest EER on every anonymization system, with particularly large margins over the best existing methods on T10-2 (7.04\% vs.\ 22.46\%) and T12-5 (18.89\% vs.\ 25.63\%). Even DAST trained with only 10\% of Stage~III data outperforms all baselines across all seven systems, further underscoring the strength of the Stage~II representations. These results again demonstrate that the proposed DAST, improved through both architecture and training strategy, is highly effective.

\section{Conclusion}

In this work, we presented DAST, a dual-stream voice anonymization attacker that jointly addresses architecture design and training strategy through a three-stage curriculum. Through systematic ablation studies, we validated that each stage contributes a distinct capability: Stage I provides the speaker-discriminative foundation, Stage II builds anonymization-invariant robustness, and Stage III sharpens target-specific decision boundaries.
A key finding is that the anonymization-invariant representations from Stage II enable highly sample-efficient adaptation against unseen anonymization systems. To this end, we publicly release the Stage II model %(\textbf{Foundation Attacker})
as a general-purpose pre-trained checkpoint, allowing practitioners to fine-tune it on any target system without repeating the costly upstream training.

\section{Acknowledgments}
This work is supported by the Singapore Ministry of Education (MOE) Academic Research Fund (AcRF) Tier 1 grant R-R13-A405-0005.

\section{Generative AI Use Disclosure}
Large language models (LLMs) were used in a limited capacity as auxiliary tools to assist with the language polishing of the manuscript. All research ideas, methodological design, experiments, analyses, and final writing decisions were conducted and verified by the authors.

\bibliographystyle{IEEEtran}
\bibliography{mybib}

\end{document}